\documentclass[journal]{IEEEtran}
\usepackage{cite}

\ifCLASSINFOpdf
 \usepackage[pdftex]{graphicx}
\else
\fi
\usepackage{amsmath}
\usepackage{algorithm}
\usepackage{algorithmicx}
\usepackage{amsmath,algorithm,tabularx}
\usepackage{algpseudocode}
\usepackage{amssymb}
\usepackage{multirow}
\usepackage{makecell}
\usepackage{caption}
\usepackage{array}
\newcolumntype{P}[1]{>{\centering\arraybackslash}p{#1}}
\newcolumntype{C}{>{\centering\arraybackslash} m{6cm} }

\usepackage{url}

\makeatletter
\newcommand{\multiline}[1]{%
	\begin{tabularx}{\dimexpr\linewidth-\ALG@thistlm}[t]{@{}X@{}}
		#1
	\end{tabularx}
}
\makeatother

\begin{document}
%
\title{Practical Scheduling Algorithms with Contiguous Resource Allocation for Next-Generation Wireless Systems}
%
%
%

\author{Shu~Sun
        and~Sungho~Moon
\thanks{The authors are with the Next Generation and Standards Group, Intel Corporation, Santa Clara, CA 95054 USA (e-mail: shu.sun@intel.com; sungho.moon@intel.com).

This work has been submitted to the IEEE for possible publication. Copyright may be transferred without notice, after which this version may no longer be accessible.}}
%
%

%

\maketitle

\begin{abstract}
This paper proposes three novel resource and user scheduling algorithms with contiguous frequency-domain resource allocation (FDRA) for wireless communications systems. The first proposed algorithm jointly schedules users and resources selected adaptively from both ends of the bandwidth part (BWP), while the second and third ones apply disjoint user and resource selection with either single-end or dual-end BWP strategies. Distinct from existing contiguous FDRA approaches, the proposed schemes comply with standards specifications for fifth-generation (5G) and beyond 5G communications, and have lower computational complexity hence are more practical. Simulation results show that all of the proposed algorithms can achieve near-optimal performance in terms of throughput and packet loss rate for low to moderate traffic load, and the first one can still perform relatively well even with a large number of users. 
\end{abstract}

\begin{IEEEkeywords}
Beyond 5G (B5G), quality of service (QoS), frequency-domain resource allocation (FDRA), scheduling.
\end{IEEEkeywords}

\IEEEpeerreviewmaketitle

\section{Introduction}
\IEEEPARstart{I}{n} a wireless communications system with a next-generation NodeB (gNB) and multiple user equipments (UEs), a pivotal issue to tackle is the scheduling of available time and frequency resources to the UEs in order to satisfy certain quality of service (QoS) requirements such as throughput, fairness, latency, and/or reliability. According to the specifications by the 3rd Generation Partnership Project (3GPP)~\cite{38214,38331}, there are two types of downlink frequency-domain resource allocation (FDRA): type 0 and type 1. In downlink FDRA of type 0, the resource block (RB) assignment information comprises a bitmap indicating the resource block groups (RBGs) that are allocated to the scheduled UE, where an RBG consists of a set of consecutive virtual RBs defined by higher layer parameters~\cite{38214}. In downlink FDRA of type 1, the RB assignment information signifies to a scheduled UE a set of contiguously allocated non-interleaved or interleaved virtual RBs within the active bandwidth part (BWP)~\cite{38214}. Two key discrepancies between type-0 and type-1 FDRA are (1) type 0 is on the RBG level while type 1 is on the RB level, and (2) the resources (RBGs or RBs) assigned to each UE can be non-contiguous for type 0, while they must be contiguous for type 1. 

A variety of scheduling methods for contiguous FDRA have been proposed previously~\cite{Wong11,Zhang13,Cicalo14,Maciel18,Arafat19,Tsiropoulou16}, predominantly for single-carrier frequency division multiple access in the Long-Term-Evolution-Advanced system. An optimal algorithm was presented in~\cite{Wong11}, which yielded the best performance but was quite intricate. To reduce the complexity, a greedy heuristic allocation was then proposed in~\cite{Wong11} which performed adjacent RB allocation expansion around a localized optimal RB for each UE. At each iteration, the UE and feasible RB combination arousing the largest increase in weighted capacity was selected. In~\cite{Zhang13}, a two-step FDRA scheme was proposed prioritizing the most demanding UEs in terms of their QoS requirements. The authors of~\cite{Cicalo14} proposed a sub-optimal algorithmic solution to address the problem of ergodic sum-rate maximization with the constraint of consecutive RB allocation, where the performance gap to optimal solution was limited to 10\%. The invention in~\cite{Maciel18} also contained two steps where the allocation leading to maximum throughput was first found without considering the contiguity constraint, which was then iteratively refined to reach an allocation satisfying contiguity. The algorithm presented in~\cite{Arafat19} was based on channel gain matrix and iterative RB cluster selection with highest channel gain. More prior art can be found in~\cite{Tsiropoulou16}. The existing strategies, however, mainly concentrate on throughput and do not take into account other QoS criteria such as delay and packet drop rate~\cite{Wong11,Zhang13,Cicalo14,Maciel18,Arafat19}, and/or involve high computational complexity when the number of UEs is large~\cite{Zhang13,Maciel18}. 
In this article, we put forth three scheduling algorithms with type-1 (i.e., contiguous) FDRA which are aligned with 3GPP standards for fifth-generation (5G) and beyond-5G (B5G) communications~\cite{38214,38331} and have relatively low computational complexity thus viable to deploy in practice. In the first proposed algorithm, FDRA and UE scheduling are jointly conducted to achieve near-optimal performance, whereas in the second and third proposed algorithms, UE selection is executed first, followed by RB allocation, whose major advantage is low complexity. Moreover, all of the proposed algorithms are flexible in terms of scheduling metric such as sum-rate, proportional fairness (PF)~\cite{Tse99} and modified largest weighted delay first (M-LWDF)~\cite{Andrews01}. System-level simulations are carried out to validate and compare the performance of the proposed algorithms, using several traffic types with diverse packet sizes, arrival rates, and QoS requirements.  

\vspace{-8pt}
\section{System Model and Problem Formulation}
\begin{figure}
	\centering
	\includegraphics[width=\columnwidth]{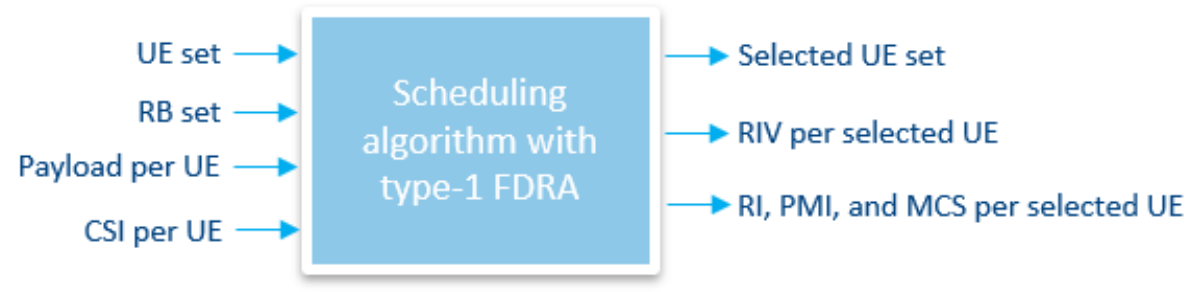}
	\caption{Input and output relation per slot in the UE and resource scheduling process for a multi-UE scenario with type-1 FDRA.}
	\label{fig:inputOutput}	
\end{figure}

In this work, we investigate a downlink cellular system comprising of one gNB and $K$ UEs indexed by the set $\mathcal{K}=\{0,...,K-1\}$, where the UEs' traffic types can be heterogenous with dissimilar QoS requirements. The transmission BWP $W$ is orthogonally divided into $B$ RBs indexed by the set $\mathcal{B}=\{0,...,B-1\}$. The payload for UE $k$ is denoted by $L_k$. There are two constraints in type-1 FDRA in 3GPP 5G and B5G specifications~\cite{38214,38331}: (1) exclusivity, meaning an RB can only be allocated to at most one UE; (2) contiguity, i.e., the RBs assigned to each UE must be contiguous.  Fig.~\ref{fig:inputOutput} illustrates the input and output relation per slot in the UE and resource scheduling process with type-1 FDRA~\cite{38214,38331}. The input incorporates the UE set $\mathcal{K}$, RB set $\mathcal{B}$, payload and channel state information (CSI) per UE, where the CSI usually embodies rank indicator (RI), precoding matrix indicator (PMI), and channel quality indicator (CQI)~\cite{38214}. The output includes the selected UE set $\mathcal{K}^\star$, selected RB set $\mathcal{B}^\star$ implied by resource indication value (RIV) per selected UE, and RI, PMI, and modulation and coding scheme (MCS) per selected UE, where an RIV corresponds to a starting virtual RB and a length pertaining to contiguously allocated RBs~\cite{38214}. 

\vspace{- 12pt} 
In some of the proposed algorithms to be elaborated in Section III, the calculation of transport block size (TBS) [1] over a certain number of RBs is needed. If wideband (WB) CSI [1] is available, where WB denotes the entire active BWP, the TBS is directly computed using the WB CSI. If subband (SB) CSI is available (while RI is still WB) [1], where an SB is equivalent to an RBG herein, the TBS over all the selected RBs (or RBGs) is calculated via the procedure below: (1) Convert each SB CQI to SB MCS, (2) compute the effective MCS over all the selected RBs (or RBGs), and (3) calculate the TBS over all the selected RBs (or RBGs) using the effective MCS and WB RI. In this article, the effective MCS equals the median of the MCSs over all the selected RBs (or RBGs), but it can also be the average, maximum, or other quantities related to SB MCS. Further, to obtain WB CQI to be utilized in some of the proposed algorithms, the effective MCS is computed over the entire active BWP, which is then converted back to CQI. 

For UE $k$ on RB $b$, given the estimated channel $\mathbf{H}_{k,b}$ and precoding matrix codebook~\cite{38214}, the RI, PMI, and MCS can be obtained via Algorithms 1 and 2 in~\cite{Sun20}, after which $\mathrm{TBS}_{k,b}$  is calculated via the method mentioned above. The achievable rate of UE $k$ on RB $b$ in each slot is $r_{k,b}=\mathrm{TBS}_{k,b}$. Let $\mathcal{B}_k$ denote the set of RBs allocated to UE $k$, the achievable rate of UE $k$ is $r_k=\sum_{b \in\mathcal{B}_k}r_{k,b}$. The scheduling metric (e.g., sum-rate, PF, M-LWDF) can be flexible depending on the system requirement. Considering QoS requirement, we select M-LWDF as the scheduling metric as an example, which is expressed as~\cite{Andrews01} $\mu_{k,b}=-(\log\delta_k/\tau_k)d_kr_{k,b}/R_k$, where $\delta_k$, $\tau_k$, $d_k$, and $R_k$ denote the acceptable packet drop probability, delay threshold (the maximum allowable delay from packet generation to packet scheduling), head-of-line (HOL) delay, and historical average rate of UE $k$, respectively. The optimization problem is formulated as  
\begin{equation}\label{eq1}
\begin{alignedat}{2}
\text{(P1):}~ &\max_{\{\mathcal{B}_0,...,\mathcal{B}_{K-1}\}\subseteq\mathcal{A}}&&\sum_{k \in\mathcal{K}}\sum_{b \in\mathcal{B}_k}\mu_{k,b} \\
&\text{subject~to} & & \mathcal{B}_k\cap\mathcal{B}_{k^\prime}=\emptyset,\forall k\ne k^\prime, k,k^\prime\in\mathcal{K}, \\
& & &d_k\leq\tau_k, \forall k\in\mathcal{K}
\end{alignedat}
\end{equation}

\noindent where $\mathcal{A}$ is the set of all possible RB allocations satisfying the contiguity constraint. (P1) is non-convex whose optimal solution requires exhaustive search with prohibitively high computational complexity. Therefore, in the next section, we propose three practical sub-optimal algorithms to tackle (P1). 

\vspace{- 9.9pt} 
\section{Scheduling Algorithms}
\subsection{Proposed Three Scheduling Algorithms}
\vspace{- 3pt} 
Since it is almost impossible to obtain the optimal solution to (P1) with reasonable computational complexity, we propose three sub-optimal algorithms to solve (P1). Inspired by the observation that $\sum_{k \in\mathcal{K}}\sum_{b \in\mathcal{B}_k}\mu_{k,b}$ in (\ref{eq1}) is likely to be maximized if the UEs who yield the largest $\sum_{b \in\mathcal{B}_k}\mu_{k,b}$ while consuming the minimum resources are scheduled first, we propose an algorithm named Joint Allocation with Dual Ends (JADE), whose procedures are detailed in Algorithm 1. Essentially, JADE jointly prioritizes the UE and RB(s) in each allocation step that produces the largest scheduling metric with the minimum number of RBs, where the RB selection is performed and compared between both ends of the active BWP to take advantage of  frequency diversity. It is worth noting that a variant of JADE, where the RBs are allocated from only one end, rather than both ends, of the BWP, can be applied as well. As the variant is likely to yield inferior performance to JADE due to less frequency diversity, its performance is not shown herein. 
\begin{algorithm}
	\caption{Joint Allocation with Dual Ends (JADE)}
	\begin{algorithmic}[1]
		\Require {Initialize $\mathcal{K}^\star=\emptyset$, $\mathcal{B}^\star=\emptyset$.}
		\While{$\mathcal{K}\neq\emptyset$ and $\mathcal{B}\neq\emptyset$} 
		\For{$\forall k\in\mathcal{K}$}
		\State \multiline{%
			Calculate the number of RBs needed, $n_{k,\text{start}}$, to transmit $L_k$ starting from the first remaining RB in $\mathcal{B}$ and going forward, until $r_{k,\text{start}}\geq L_k$ or $\mathcal{B}=\emptyset$. Denote the selected RB set as $\mathcal{B}_{k,\text{start}}$.}
		\State \multiline{%
			Calculate the number of RBs needed, $n_{k,\text{end}}$, to transmit $L_k$ starting from the last remaining RB in $\mathcal{B}$ and going backward, until $r_{k,\text{end}}\geq L_k$ or $\mathcal{B}=\emptyset$. Denote the selected RB set as $\mathcal{B}_{k,\text{end}}$.}
		\State \multiline{%
			If $n_{k,\text{start}}\leq n_{k,\text{end}}$, store $\mathcal{B}_{k,\text{start}}$ and $r_{k,\text{start}}$ as $\mathcal{B}_k$ and $r_k$, respectively; otherwise store $\mathcal{B}_{k,\text{end}}$ and $r_{k,\text{end}}$ as $\mathcal{B}_k$ and $r_k$, respectively.}
		\State \multiline{%
			Calculate $\sum_{b \in\mathcal{B}_k}\mu_{k,b} $.}
		\EndFor
		\State \multiline{%
			$k^\star=\underset{k}{\mathrm{argmax}}\sum_{b \in\mathcal{B}_k}\mu_{k,b}$.}
		\State \multiline{%
			Calculate $\text{MCS}_{k^\star}$, the final MCS for UE $k^\star$ over $\mathcal{B}_{k^\star}$.}
		\State \multiline{%
			$\mathcal{K}^\star\gets\mathcal{K}^\star\cup\{k^\star\}$, $\mathcal{B}^\star\gets\mathcal{B}^\star\cup\mathcal{B}_{k^\star}$.}
		\State \multiline{%
			$\mathcal{K}\gets\mathcal{K}\setminus\{k^\star\}$, $\mathcal{B}\gets\mathcal{B}\setminus\mathcal{B}_{k^\star}$.}
		\EndWhile
		\State \textbf{return} $\mathcal{K}^\star$, $\mathcal{B}^\star$, and $\text{MCS}_k,\forall k\in\mathcal{K}^\star$. 
	\end{algorithmic}
\end{algorithm}

Note that in JADE, the number of TBS and scheduling metric calculation is proportional to $K^2$ due to the iteration for each remaining UE and RB. To further reduce the computational complexity, two lower-complexity algorithms are designed, i.e., Disjoint Allocation with Single End (DASE) and Disjoint Allocation with Two Ends (DATE). The main design principle of DASE and DATE is to guarantee the QoS for UEs with the most stringent delay and acceptable packet drop probability requirements. In both DASE and DATE, UE selection is done first based on their delay and acceptable packet drop probability requirements as well as the number of RBs needed, followed by RB selection. The only difference between DASE and DATE lies in that RB selection is conducted from only one end of the BWP in DASE, while both ends of the BWP are considered and compared for RB selection in DATE. Detailed steps for DASE and DATE are provided in Algorithm 2 and Algorithm 3, respectively. 
\vspace{- 5pt} 
\begin{algorithm}
	\caption{Disjoint Allocation with Single End (DASE)}
	\begin{algorithmic}[1]
		\Require {Initialize $\mathcal{K}^\star=\emptyset$, $\mathcal{B}^\star=\emptyset$.}
		\While{$\mathcal{K}\neq\emptyset$ and $\mathcal{B}\neq\emptyset$} 
		\For{$\forall k\in\mathcal{K}$} \Comment{UE selection begins}
		\State \multiline{%
			$\Delta d_k=\tau_k-d_k$.}
		\State \multiline{%
			Calculate the number of RBs needed, $n_k$, to transmit $L_k$ based on WB CQI of UE $k$.}
		\EndFor
		\State \multiline{%
			$\mathcal{K}^\star_{\text{temp}}=\underset{k}{\mathrm{argmin}}\Delta d_k$.}
		\If{$|\mathcal{K}^\star_{\text{temp}}|=1$}
		\State \multiline{%
			$k^\star=\underset{k}{\mathrm{argmin}}~\Delta d_k$.}
		\Else
		\State \multiline{%
			$\mathcal{K}^\star_{\text{temp}}=\underset{k}{\mathrm{argmin}}~\delta_k$.}
		\If{$|\mathcal{K}^\star_{\text{temp}}|=1$}
		\State \multiline{%
			$k^\star=\underset{k}{\mathrm{argmin}}~\delta_k$.}
		\Else
		\State \multiline{%
			$\mathcal{K}^\star_{\text{temp}}=\underset{k}{\mathrm{argmin}}~n_k$.}
		\If{$|\mathcal{K}^\star_{\text{temp}}|=1$}
		\State \multiline{%
			$k^\star=\underset{k}{\mathrm{argmin}}~n_k$.}
		\Else
		\State \multiline{%
			Randomly select a UE $k^\star$.}
		\EndIf
		\EndIf
		\EndIf \Comment{UE selection ends and RB selection begins}
		\State \multiline{%
			Calculate the number of RBs needed, $n_{k^\star,\text{start}}$, to transmit $L_{k^\star}$ starting from the first remaining RB in $\mathcal{B}$ and going forward, until $r_{k^\star,\text{start}}\geq L_{k^\star}$ or $\mathcal{B}=\emptyset$. Denote the selected RB set as $\mathcal{B}_{k^\star}$, and allocate $\mathcal{B}_{k^\star}$ to UE $k^\star$.} \Comment{RB selection ends}
		\State \multiline{%
			Calculate $\text{MCS}_{k^\star}$, the final MCS for UE $k^\star$ over $\mathcal{B}_{k^\star}$.}
		\State \multiline{%
			$\mathcal{K}^\star\gets\mathcal{K}^\star\cup\{k^\star\}$, $\mathcal{B}^\star\gets\mathcal{B}^\star\cup\mathcal{B}_{k^\star}$.}
		\State \multiline{%
			$\mathcal{K}\gets\mathcal{K}\setminus\{k^\star\}$, $\mathcal{B}\gets\mathcal{B}\setminus\mathcal{B}_{k^\star}$.}
		\EndWhile
		\State \textbf{return} $\mathcal{K}^\star$, $\mathcal{B}^\star$, and $\text{MCS}_k,\forall k\in\mathcal{K}^\star$. 
	\end{algorithmic}
\end{algorithm}

\begin{algorithm}
	\caption{Disjoint Allocation with Two Ends (DATE)}
	\begin{algorithmic}[1]
		\Require {Initialize $\mathcal{K}^\star=\emptyset$, $\mathcal{B}^\star=\emptyset$.}
		\While{$\mathcal{K}\neq\emptyset$ and $\mathcal{B}\neq\emptyset$} 
		\State \multiline{%
			UE selection: Identical to Steps 2-21 in DASE.}
		\Comment{RB selection begins}
		\State \multiline{%
			Calculate the number of RBs needed, $n_{k^\star,\text{start}}$, to transmit $L_{k^\star}$ starting from the first remaining RB in $\mathcal{B}$ and going forward, until $r_{k^\star,\text{start}}\geq L_{k^\star}$ or $\mathcal{B}=\emptyset$. Denote the selected RB set as $\mathcal{B}_{k^\star,\text{start}}$.} 
		\State \multiline{%
			Calculate the number of RBs needed, $n_{k^\star,\text{end}}$, to transmit $L_{k^\star}$ starting from the last remaining RB in $\mathcal{B}$ and going backward, until $r_{k^\star,\text{end}}\geq L_{k^\star}$ or $\mathcal{B}=\emptyset$. Denote the selected RB set as $\mathcal{B}_{k^\star,\text{end}}$.}
		\State \multiline{%
			If $n_{k^\star,\text{start}}\leq n_{k^\star,\text{end}}$, denote $\mathcal{B}_{k^\star,\text{start}}$ as $\mathcal{B}_{k^\star}$, otherwise denote $\mathcal{B}_{k^\star,\text{end}}$ as $\mathcal{B}_{k^\star}$. Allocate $\mathcal{B}_{k^\star}$ to UE $k^\star$.} \Comment{RB selection ends}
		\State \multiline{%
			Calculate $\text{MCS}_{k^\star}$, the final MCS for UE $k^\star$ over $\mathcal{B}_{k^\star}$.}
		\State \multiline{%
			$\mathcal{K}^\star\gets\mathcal{K}^\star\cup\{k^\star\}$, $\mathcal{B}^\star\gets\mathcal{B}^\star\cup\mathcal{B}_{k^\star}$.}
		\State \multiline{%
			$\mathcal{K}\gets\mathcal{K}\setminus\{k^\star\}$, $\mathcal{B}\gets\mathcal{B}\setminus\mathcal{B}_{k^\star}$.}
		\EndWhile
		\State \textbf{return} $\mathcal{K}^\star$, $\mathcal{B}^\star$, and $\text{MCS}_k,\forall k\in\mathcal{K}^\star$.   
	\end{algorithmic}
\end{algorithm}

\subsection{Scheduling Algorithm with Type-0 FDRA}
Ideally, the performance of the proposed algorithms should be compared with that of the optimal type-1 FDRA which, however, requires exhaustive search over UEs, the starting position of RBs per UE, and the number of RBs per UE, whose complexity is prohibitively high and hence almost impossible to realize. On the other hand, although also requiring exhaustive search, the optimal type-0 FDRA is possible to realize capitalizing on a different and smaller search space. Additionally, if the same frequency granularity is assumed for both type-0 and type-1 FDRA, the optimal type-0 allocation is expected to be superior to the optimal type 1, since discontinuous FDRA enjoys higher flexibility in terms of best UE and resource combination selection. To this end, we compare the performance of the proposed scheduling algorithms with one using the optimal type-0 FDRA which serves as a benchmark. In type-0 FDRA, the scheduling metric is first calculated per UE per RBG, then the UE and RBG combination corresponding to the largest scheduling metric is selected for scheduling and excluded from further selection afterwards in the current slot. Subsequently, the scheduling metric is recalculated per UE per RBG, followed by best UE and RBG combination selection and exclusion, so on and so forth, until there is no remaining UE or RBG. 

\vspace{- 10pt} 
\subsection{Complexity Analysis}
Besides the RBG-level type-0 FDRA, we also compare the proposed algorithms with a representative sub-optimal contiguous FDRA algorithm in the industry published in~\cite{Wong11}, which is named Localized Expansion of Adjacent Positions (LEAP) herein, to evaluate the performance enhancement by the proposed algorithms over LEAP. The weighted capacity metric in LEAP is replaced by the scheduling metric for fair comparison. Assuming each UE needs $M$ RBs on average, and the number of RBs in an RBG is $M_\text{RB}$, the computational complexity of all the considered algorithms is provided in Table~\ref{tbl:complexity}. As expected, type-0 FDRA possesses the highest complexity due to exhaustive search over all UE and RBG combinations. In a typical example where $K=30, B=270,M=10$, and $M_\text{RB}=4$, the complexity of JADE is slightly lower than that of LEAP, both of which are approximately an order of magnitude lower than type-0 FDRA. DASE has the lowest complexity which is slightly lower than that of DATE, since these two algorithms exploit disjoint UE and RB selection.  
\begin{table}[!t]
	\renewcommand{\arraystretch}{1.1}
	\caption{Complexity Comparison}
	\label{tbl:complexity}
	\centering
	\begin{tabular}{|m{1.7cm}||m{0.75cm}|m{0.75cm}|m{0.75cm}|m{0.75cm}|m{1.1cm}|}
		\hline
		\multirow{2}{*}{Algorithm} & \multicolumn{4}{c|}{Type-1} & \multirow{2}{*}{Type-0}\\
		\cline{2-5}
		& JADE & DASE & DATE & LEAP &\\
		\hline
		\makecell{Number of \\TBS calculation} & $MK^2$ & $MK$ & $2MK$ & $MK$ & 0\\
		\hline
		\makecell{Number of\\ scheduling \\metric\\calculation} & $\frac{MK^2}{2}$ & 0 & 0 & $\frac{BK^2}{3}$ & $\frac{[\frac{M}{M_\text{RB}}]^2K^3}{3}$\\
		\hline
		\makecell{Number \\of RB amount \\calculation} & 0 & $K$ & $K$ & 0 & 0\\
		\hline
		\makecell{Sum\\complexity} & $\frac{3MK^2}{2}$ & $MK$ & $2MK$ & $\frac{BK^2}{3}$ & $\frac{[\frac{M}{M_\text{RB}}]^2K^3}{3}$\\
		\hline
		\makecell{Sum\\complexity for\\$K=30$, \\$B=270$,\\$M=10$, \\$M_\text{RB}=4$} & \makecell{$1e4$\\$\Downarrow$\\$\mathcal{O}(1e4)$} & \makecell{$3e2$\\$\Downarrow$\\$\mathcal{O}(1e2)$}  & \makecell{$6e2$\\$\Downarrow$\\$\mathcal{O}(1e2)$}  & \makecell{$8e4$\\$\Downarrow$\\$\mathcal{O}(1e4)$}  & \makecell{$2e5$\\$\Downarrow$\\$\mathcal{O}(1e5)$} \\
		\hline
	\end{tabular}
\end{table}

\vspace{- 9pt} 
\section{Simulation Results}
System-level simulations are conducted to assess and compare the performance of the proposed three type-1 FDRA algorithms. Table~\ref{tbl:simSet} lists the simulation settings, where the traffic models comprise both eMBB (enhanced Mobile Broadband) and URLLC (Ultra-Reliable Low-Latency Communications)~\cite{Bennis18} (including arVr2, powerDist2, and ITS~\cite{38824}, where arVr2 denotes the second type of augmented reality/virtual reality, powerDist2 represents the second type of power distribution grid fault and outage management, and ITS stands for intelligent transportation system~\cite{38824}). The total number of UEs in our simulations vary from 10 to 50, and the ratios of eMBB, arVr2, powerDist2, and ITS UEs are about 1:1:1:1. Table~\ref{tbl:traffic} details the parameters for the traffic models studied in our simulations. We note that URLLC traffic can also be scheduled by puncturing the ongoing eMBB transmission, but that scheme has its own drawbacks as well and is out of the scope of this paper whose overarching focus is the scheduling of different traffic types using the same time resource. 
\begin{table}[!t]
	\renewcommand{\arraystretch}{1.0}
	\caption{Simulation Settings}
	\label{tbl:simSet}
	\centering
	\begin{tabular}{|c||c|}
		\hline
		Configuration & Value \\
		\hline
		Transmit power & 23 dBm \\
		\hline
		Number of gNB antennas & 4 \\
		\hline
		Cell radius & 250 m \\
		\hline
		UE distribution & Uniform \\
		\hline
		Number of antennas per UE & 4 \\
		\hline
		Number of UEs per gNB & 10-50 \\
		\hline
		Channel& \makecell{EPA20 (6.0 km/h) (Extended \\Pedestrian A model with 20 Hz \\Doppler frequency)} \\
		\hline
		Numerology& \makecell{30kHz sub-carrier spacing, \\100MHz bandwidth} \\
		\hline
		CSI feedback delay & 1 slot \\
		\hline
		Traffic model& \makecell{eMBB, arVr2, ITS, powerDist2} \\
		\hline
		Traffic ratio& \makecell{1:1:1:1} \\
		\hline
		Number of slots& \makecell{1200 per seed} \\
		\hline
		\makecell{Number of seeds \\per simulation run}& \makecell{10} \\
		\hline
		\makecell{Number of RBs per RBG}& \makecell{4} \\
		\hline
	\end{tabular}
\end{table}

\begin{table}[!t]
	\renewcommand{\arraystretch}{1.1}
	\caption{Parameters for Traffic Models used in Simulations~\cite{38824}}
	\label{tbl:traffic}
	\centering
	\begin{tabular}{|c||c|c|c|c|}
		\hline
		 & eMBB & arVr2 & ITS & powerDist2 \\
		\hline
		\makecell{Delay threshold (ms)}& 100 & 7 & 7 & 6\\
		\hline
		\makecell{Acceptable packet \\drop probability}& 10\%	&0.1\%	&0.001\%	&0.001\%\\
		\hline
		\makecell{Packet size (bits)}& 12000 &32768	&10960	&2000\\
		\hline
		\makecell{Packet arrival rate \\(packets/second)}& 1000	&60	&100 &1200\\
		\hline
	\end{tabular}
\end{table}
\setlength{\belowcaptionskip}{-10pt}

\begin{figure}
	\centering
	\includegraphics[width=\columnwidth]{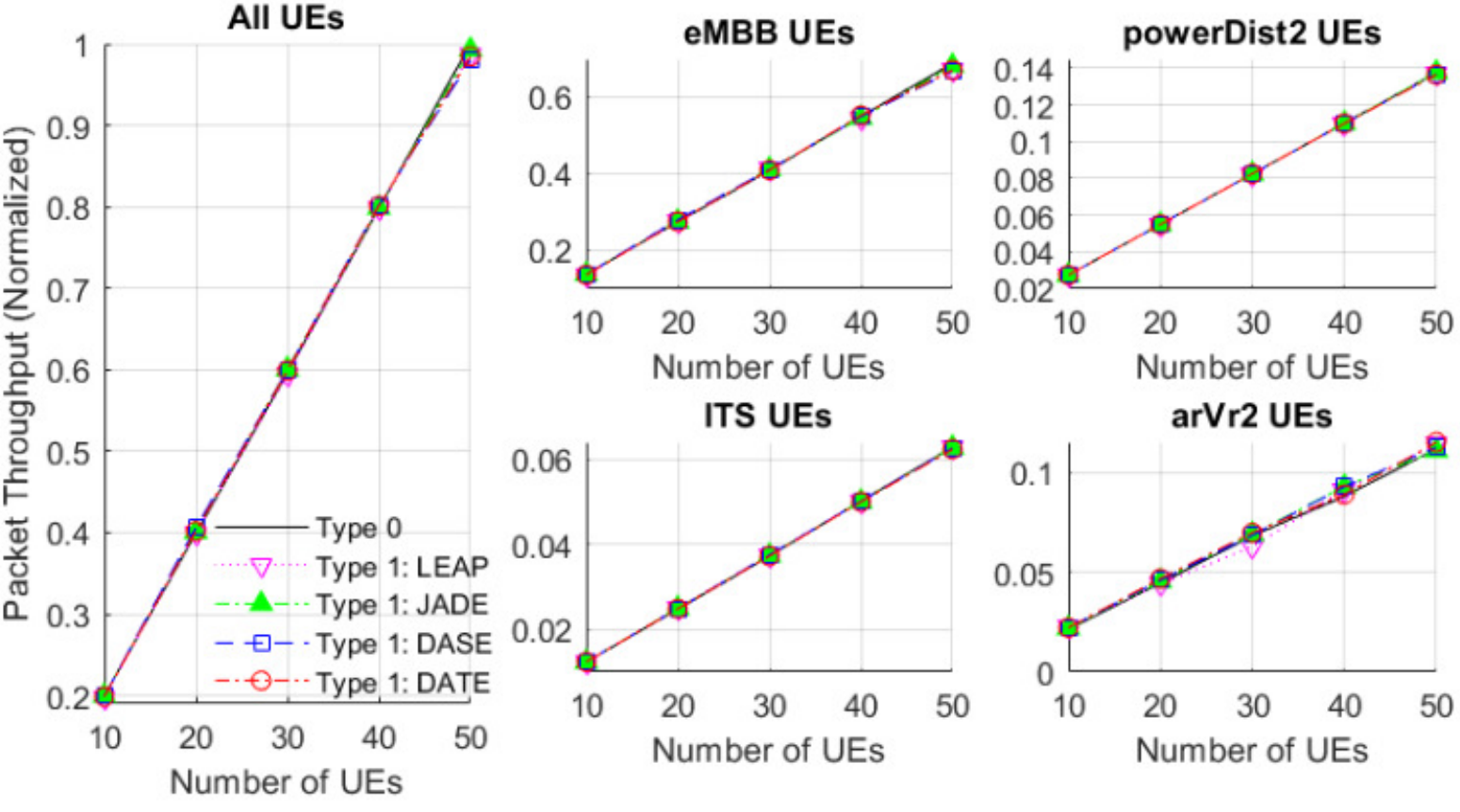}
	\caption{Packet throughput of the proposed three scheduling algorithms with type-1 FDRA, as well as LEAP in~\cite{Wong11} and an RBG-level type-0 FDRA.}
	\label{fig:tputAll}	
\end{figure}

\begin{figure}
	\centering
	\includegraphics[width=\columnwidth]{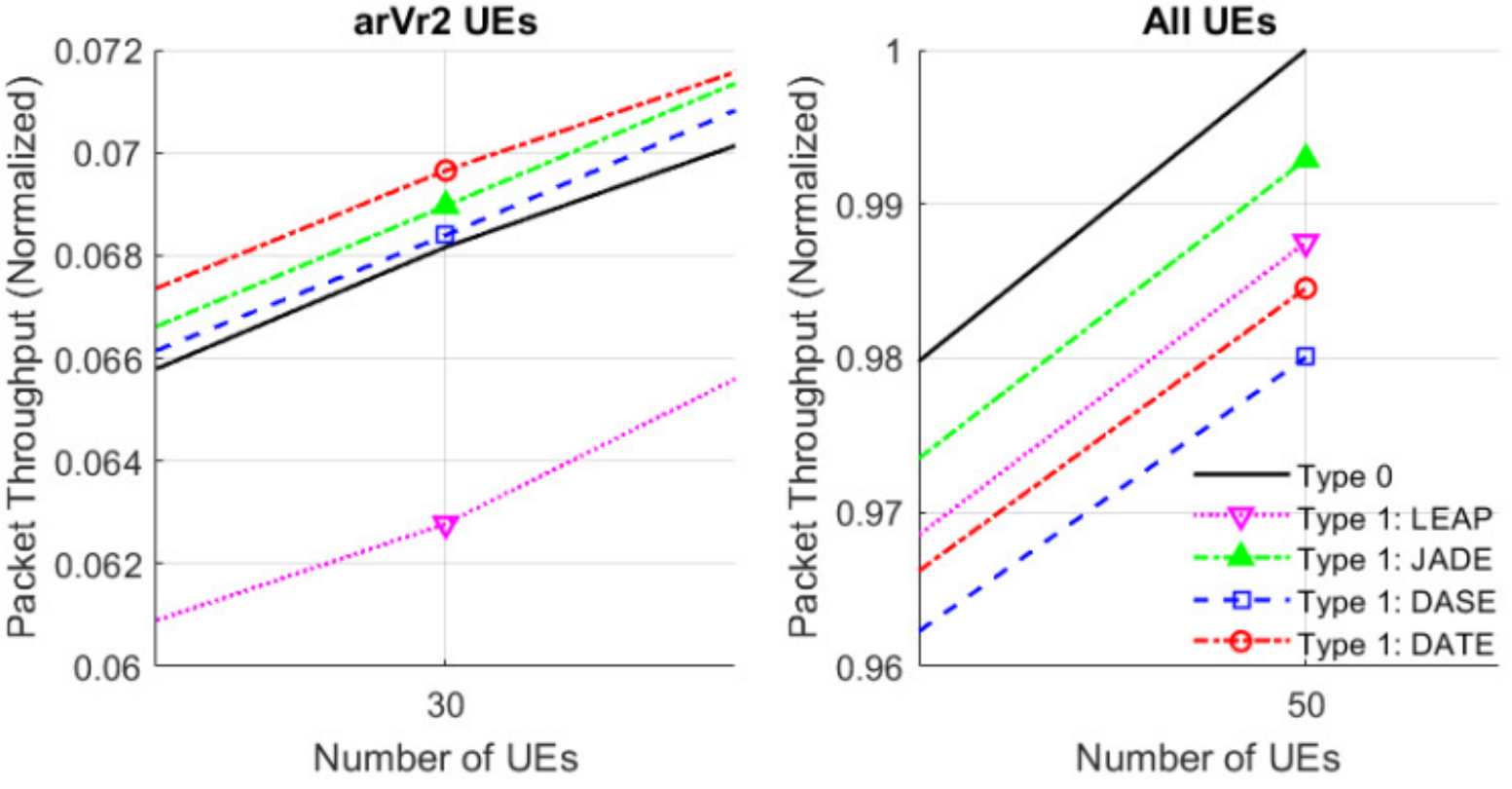}
	\caption{Close-up views of packet throughput for arVr2 UEs in the case of 30 total UEs and for all UEs in the case of 50 total UEs.}
	\label{fig:tput}	
\end{figure}

\begin{figure}
	\centering
	\includegraphics[width=0.81\columnwidth]{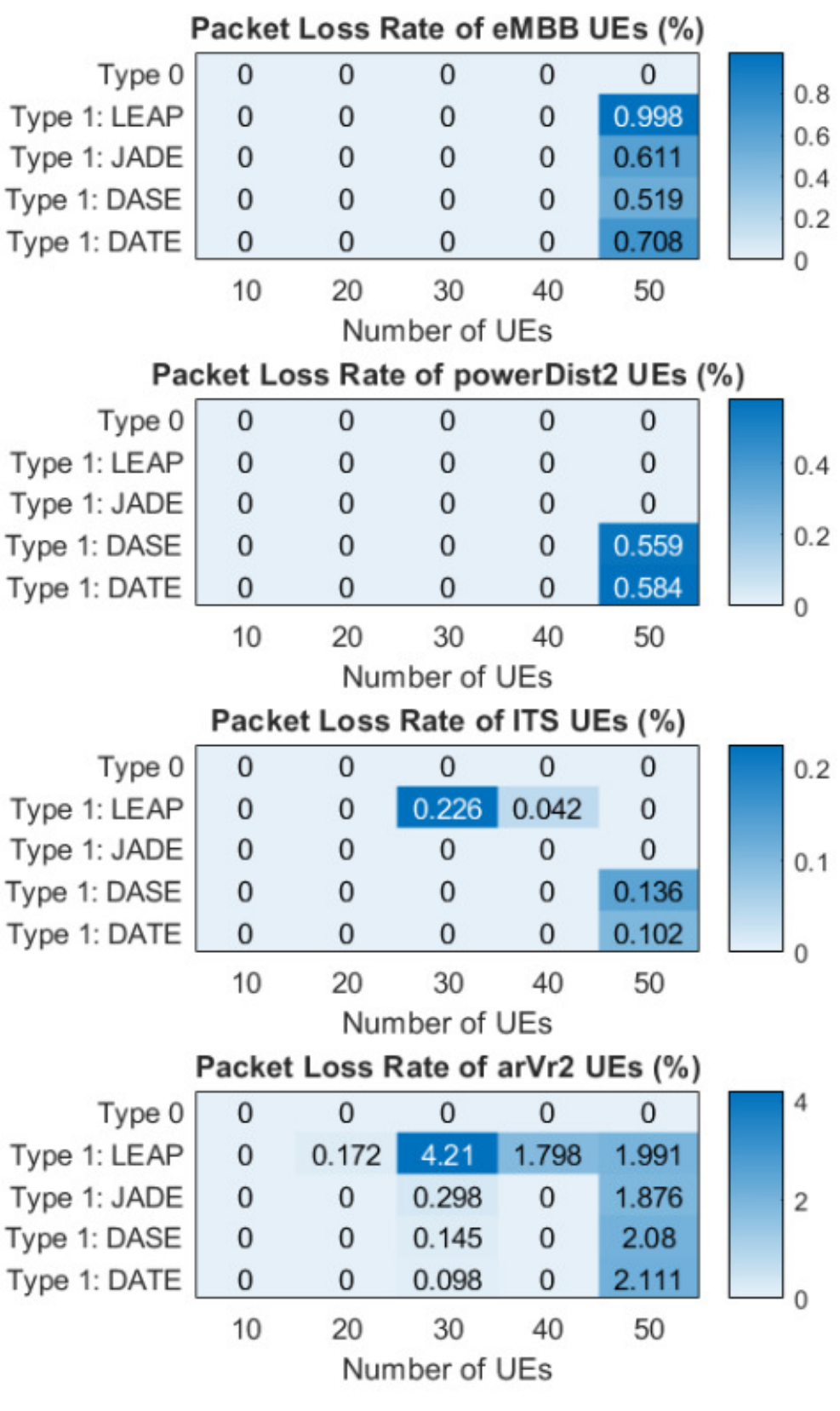}
	\caption{Packet loss rate of the proposed three scheduling algorithms with type-1 FDRA, as well as LEAP in~\cite{Wong11} and an RBG-level type-0 FDRA. }
	\label{fig:pkLoss}	
\end{figure}

The overall and two close-up views of packet throughput for various traffic types are illustrated in Fig.~\ref{fig:tputAll} and Fig.~\ref{fig:tput}, respectively. The maximum throughput degradation of the proposed algorithms against the type-0 algorithm and maximum throughput gain over LEAP are summarized in Table~\ref{tbl:tput}. Fig.~\ref{fig:pkLoss} shows the packet loss rate, where a packet is considered lost if it is not entirely scheduled before reaching its delay threshold. For LEAP, the swift increase in packet loss at 30 UEs in Fig.~\ref{fig:pkLoss} for ITS and arVr2 unveils its instability and sensitivity to the UE amount and/or locations. The following key observations can be drawn from these results: 

\noindent 1) In general, JADE outperforms DASE, DATE as well as LEAP. The superiority of JADE over DASE and DATE is especially evident when the number of UEs is large (e.g., 50), as shown by the throughput and packet loss rate for 50 UEs in Fig.~\ref{fig:tputAll} to Fig.~\ref{fig:pkLoss}. Comparing JADE and LEAP, as shown in Fig.~\ref{fig:tput} and Fig.~\ref{fig:pkLoss}, JADE yields higher throughput and lower packet loss rate in most cases, with a maximum throughput gain of 9.9\% (see Table~\ref{tbl:tput}). The reason is that LEAP allocates RBs locally around the first best RB for each UE, but if the channel quality happens to change abruptly around the first RB, the overall channel quality may degrade hence incurring performance loss. 

\noindent 2) JADE has comparable and sometimes even better performance in contrast with the type-0 algorithm, as shown by the throughput for arVr2 UEs in Fig.~\ref{fig:tput}, since type-0 FDRA is based on the RBG level (4 RBs per RBG in the simulation) while type 1 is of RB-level which has a finer frequency granularity hence higher flexibility. The maximum throughput degradation of JADE against type-0 FDRA is only 0.9\%. 

\noindent 3) Dual-end FDRA surpasses its single-end counterpart, as validated by DATE and DASE, due to its higher frequency diversity.

\noindent 4) As demonstrated by the left plot of Fig.~\ref{fig:tput}, Table IV, and Fig.~\ref{fig:pkLoss}, DASE and DATE perform well and even exceed LEAP when the number of UEs is not very large, i.e., up to 40 UEs in this case. Since these two algorithms enjoy the least complexity, they can be used when traffic load is not too high.
\begin{table}
	\renewcommand{\arraystretch}{1.1}
	\caption{Throughput Comparison}
	\label{tbl:tput}
	\centering
	\begin{tabular}{|c||c|c|c|}
		\hline
		Algorithm & JADE & DASE & DATE \\
		\hline
		\makecell{Maximum throughput \\degradation against Type 0} & 0.9\%	&2.8\%	&2.5\% \\
		\hline
		\makecell{Maximum throughput gain over LEAP} & 9.9\%	&8.9\%	&11.0\% \\
		\hline
	\end{tabular}
\end{table}


\vspace{- 15pt} 
\section{Conclusion}
We have proposed three practical multi-UE scheduling algorithms with type-1 FDRA, i.e., JADE, DASE, and DATE, and compared their performance with each other and with an optimal non-contiguous RBG-level FDRA method and a typical contiguous FDRA algorithm LEAP in the industry. Numerical results demonstrate that JADE can achieve near-optimal performance and outperform LEAP in terms of QoS requirements while having low computational complexity. Additionally, DASE and DATE perform similarly to JADE for small to moderate numbers of UEs, but with substantially lower computational complexity, thus can be adopted in practice under low traffic load conditions. The proposed algorithms are applicable to both downlink and uplink. This work can be extended by considering coarser frequency granularities in contiguous FDRA to mitigate RIV overhead~\cite{Kittichokechai20}.

\ifCLASSOPTIONcaptionsoff
  \newpage
\fi
\vspace{- 6pt}
\bibliographystyle{IEEEtran}
\bibliography{Type-1_FDRA}

\end{document}